\documentclass{aa}  
\usepackage{graphicx}
\usepackage{txfonts}
\usepackage{hyperref}
\usepackage{xspace}

\newcommand{\erosita}{{eROSITA}\xspace}

\newcommand{\msun}{M$_{\odot\,}$}

\newcommand{\Fig}{Fig.}
\newcommand{\Figs}{Figs.}

\newcommand{\Sect}{Section}

\newcommand{\Tab}{Table}

\begin{document} 

\title{Hoinga: A supernova remnant discovered in the SRG/eROSITA All-Sky Survey eRASS1}
\titlerunning{Hoinga - A SNR discovered in eRASS1}

\author{W. Becker\inst{1,2}\thanks{E-mail: web@mpe.mpg.de}, N. Hurley-Walker\inst{3}, 
Ch. Weinberger\inst{1}, L. Nicastro \inst{4}, M.~G.~F.~Mayer \inst{1}, A. Merloni \inst{1}, 
J. Sanders \inst{1} }

%Werner Becker  \orcid{0000-0003-1173-6964}
%Luciano Nicastro \orcid{0000-0001-8534-6788}
%Martin G. Mayer \orcid{0000-0002-9771-9841}

\authorrunning{W.Becker, et al.}
 
\institute{
     Max-Planck-Institut f\"ur extraterrestrische Physik,
     Giessenbachstra{\ss}e, 85748 Garching, Germany\\
\and Max-Planck-Institut f\"ur Radioastronomie,
     Auf dem H\"ugel 69, 53121 Bonn, Germany \\ 
\and International Centre for Radio Astronomy Research,
     Curtin University, Bentley WA 6102, Australia\\
\and INAF -- Osservatorio di Astrofisica e Scienza dello Spazio di Bologna, 
     Via Piero Gobetti 93/3, I-40129 Bologna, Italy}

\date{Received 17.12.2020; accepted 12.02.2021}

 \abstract{
 {Supernova remnants (SNRs) are observable for about $(6 - 15) \times 10^4$ years before 
 they fade into the Galactic interstellar medium. With a Galactic supernova rate 
 of approximately two per century, we can expect to have of the order of 1200 SNRs 
 in our Galaxy. However, only about 300 of them are known to date, with the majority 
 having been discovered in Galactic plane radio surveys. Given that these SNRs represent 
 the brightest tail of the distribution and are mostly located close to the plane, they are 
 not representative of the complete sample. The launch of the Russian-German observatory 
 SRG/eROSITA in July 2019 brought a promising new opportunity to explore the universe.}
 {Here we report findings from the search for new SNRs in the eROSITA all-sky survey data 
 which led to the detection of one of the largest SNRs discovered at wavelengths other
 than the radio: G249.5+24.5. This source is located at a relatively high Galactic latitude, where 
 SNRs are not usually expected to be found.}
 {The remnant, `Hoinga', has a diameter of about $4\fdg4$' and  shows a 
 circular shaped morphology with diffuse X-ray emission filling almost the entire 
 remnant. Spectral analysis of the remnant emission reveals that an APEC spectrum from 
 collisionally ionised diffuse gas and a plane-parallel shock plasma model with 
 non-equilibrium ionisation are both able to provide an adequate description 
 of the data, suggesting a gas temperature of the order of kT = $0.1^{+ 0.02}_{-0.02}$ 
 keV and an absorbing column density of $N_{\rm{H}}=3.6^{+ 0.7}_{-0.6} \times 10^{20} 
 \rm{cm}^{-2}$. Various X-ray point sources are found to be located within the remnant 
 boundary but none seem to be associated with the remnant itself. 
 Subsequent searches for a radio counterpart of the Hoinga remnant identified  
 its radio emission in archival data from the Continuum HI Parkes All-Sky Survey (CHIPASS) 
 and the 408-MHz `Haslam' all-sky survey. The radio spectral index $\alpha=-0.69\pm0.08$ 
 obtained from these data definitely confirms the SNR nature of Hoinga. 
 We also analysed INTEGRAL SPI data for fingerprints of $^{44}\rm Ti$ emission, which is an ideal 
 candidate with which to study nucleosynthesis imprinting in young SNRs. Although no  $^{44}\rm Ti$ emission 
 from Hoinga was detected, we were able to set a $3\sigma$ upper flux limit of 
 $9.2 \times 10^{-5}$\,ph\,cm$^{-2}$\,s$^{-1}$.}
 {From its size and X-ray and radio spectral properties we conclude that Hoinga is a 
 middle-aged Vela-like SNR located at a distance of about twice that of the 
 Vela SNR, i.e.~at $\sim 500$ pc.}
}
\keywords{Stars: supernovae: general -- Stars: 
supernovae: individual: G249.5+24.5 -- Hoinga ISM: supernova remnants}

\maketitle
%
%-------------------------------------------------------------------

\section{Introduction}

A long series of observations have taught astronomers that there are many different types of stars. 
Findings in atomic and nuclear physics have made it possible to understand the development of these 
stars over the past few decades. According to this, the fate of a star at the end of its thermonuclear 
evolution essentially depends on only one parameter: the mass of a star decides whether its death is 
gentle or violent.  More massive stars with M $\ge 8$\msun end their lives with a supernova (SN) 
explosion, which is not only often associated with the formation of other exotic star types such as 
neutron stars (NSs) or black holes, but also represents a new beginning of stellar evolution by 
enrichment and decompression of the surrounding interstellar medium. A prominent example for this 
is the Solar System itself which shows imprints in metal abundance of a past SN which took place 
4.567 Gyr ago \citep{2012ApJ...745...22G}. 

Supernovae are considered to be rare events which happen in our Milky Way on average every $30-50$ years 
\citep[e.g.][]{2008MNRAS.391.2009K}, though no SN event has been directly observed in our Galaxy 
in the past 400 years. Indeed, in the past two millennia, only seven Galactic SN are the subject of
historical records: SN 185 (RCW 86), SN 386 (G11.2--0.3), SN 1006, SN 1054 (Crab), SN 1181 (3C58), 
SN 1572 (Tycho), and SN 1604 (Kepler); see also \cite{2017hsn..book...49S} and references therein. 
However, there are additional promising candidates discussed in the literature, such as for 
example~CAS A \citep{2017hsn..book..179G} and Vela-Jr\citep{1998Natur.396..141A}. 

Certainly, visible-band extinction of the SN emission and its distance to earth plays a 
crucial role when it comes to recognising a SN with the naked eye. A prominent example of this 
effect is demonstrated by the missing reports of the CAS A SN event which is believed to have taken place 
about 300 years ago. No widespread reports of CAS A exist in the literature of the 17th century 
\citep[cf.][]{1997NuPhA.621...83H}. A more recent example of an unrecognised SN is that of the 
youngest SN known in our Galaxy, G1.9+0.3, which was completely missed by optical 
observatories about 100 years ago \citep{2008ApJ...680L..41R}. 

In contrast to SNe which are only observable on a timescale of months to years, their remnants 
(SNRs) are detectable over a large range of the electromagnetic bands for more than 60\,000 to 
150\,000 years. However, today only about 300 SNRe are known  
\citep[cf.][]{2019JApA...40...36G}, most of which were discovered in Galactic plane 
radio surveys. Assuming that the radio lifetime of a SNR bright enough to be detected 
with current radio telescopes is at least about 60 kyr \citep[][]{1994ApJ...437..781F}, 
there is a discrepancy by a factor of between four and six between the observed and expected number of SNRs.
Even if one takes into account the fact that very massive stars may form a black hole without a luminous 
SN \citep[e.g.][]{2008ApJ...684.1336K, 2017MNRAS.468.4968A} there is still a significant mismatch
between the expected and known number of SNRs. The discrepancy is possibly explained 
by the fact that the radio sample of SNRs is not complete. Reasons that may prevent a 
radio bright remnant from being detected in radio surveys are various:

\begin{itemize}
\item A SN shock wave may expand within the hot phase of the ISM and reach a very large 
diameter until it has swept up sufficient mass from the low-density gas to form a radio 
shell. Density inhomogeneities in such a large volume will cause distortions in the shell 
and can make the identification as a SNR rather difficult, in particular in the presence 
of confusing unrelated emission from other nearby sources in the same region of the sky. 

\item A SN shock wave may expand in a very dense medium, making the SNR lifetime rather 
short, because material is quickly swept up and decelerated. Such an environment is likely 
to be relevant for example~for massive star members of OB-associations that are surrounded 
by dense molecular clouds and warm gas. Even during their short lifetime, such events 
are difficult to identify within the strong thermal radio emission from those regions.

\item There is a strong bias towards bright resolved objects in observations towards 
the inner Galaxy. 

\item Low-surface-brightness SNRs are easily missed in radio surveys if they are below 
the sensitivity limits of the surveys or if they are confused with other objects in the 
same area.

\item Old SNRs which are in the phase of dissolving into the ISM may have 
incomplete radio shells that may then prevent these sources from being identified as SNRs.

\item SNRs located away from the Galactic plane are easily missed in radio surveys, 
as this area is where these events are typically targeted.
\end{itemize}

Given these selection effects in radio surveys and the detection of unknown SNRs in previous X-ray 
surveys \citep[e.g.][]{1996rftu.proc..267P, 1996rftu.proc..239B, 1994A&A...284..573A, 
1996rftu.proc..233A,1996rftu.proc..247E,  1996rftu.proc..253F}, as well as the detection of more 
than 70 highly significant SNR candidates in our analysis of the ROSAT All-Sky-Survey data, it 
was deemed worthy to start searching for undiscovered SNRs in the first \erosita All-Sky 
Survey RASS1 \citep{Predehl2020a}. 

In this paper we report the discovery of the SNR G249.5+24.5 in the eROSITA data. With a diameter of about 
$4\fdg4$ it is among the largest SNRs discovered at wavelengths other than the radio. 
The structure of the paper is as follows: eROSITA and ROSAT observations of the remnant along with 
the data analysis are described in \S2. In \S3 we describe the analysis of archival radio data taken 
from the Continuum HI Parkes All-Sky Survey (CHIPASS) and the 408 MHz all-sky continuum survey 
in order to search for a radio counterpart to the remnant and to explore its radio emission
properties. \S4 presents our exploration of spectrometer data from INTEGRAL in order to search 
for possible ${}^{44}$Ti decay radiation associated with the remnant. In \S5 we summarise 
and discuss our results.  

\section{X-ray observations and data analysis \label{xray}}

\subsection{Hoinga in the eROSITA All-Sky Survey\label{eRASS}}
The German-built X-ray telescope eROSITA (extended R\"ontgen Survey Imaging Telescope Array) is 
one of two instruments on the Russian-German observatory SRG (Spectrum R\"ontgen-Gamma; 
Sunyaev et al., 2020). eROSITA consists of seven aligned X-ray telescopes (TM1$-$TM7), each nested with 54 
gold-coated mirror shells which have a focal length of 1600 mm. All telescopes observe 
the same sky-region simultaneously in the 0.2--8 keV band-pass though each focuses the 
collected X-rays on its own pn-CCD camera \citep{2014SPIE.9144E..1WM}. The latter is an 
improved version of the pn-CCD camera aboard XMM-Newton \citep{2001Strueder}. eROSITA has 
a spectral resolution of $\sim{70}$ eV  at 1 keV and a temporal resolution of 50 ms. Its field 
of view (FOV) is 1\degr. The on-axis effective area of all seven telescopes combined is 
slightly higher than that of the XMM-Newton pn + MOS cameras in the key 0.5-2.0 keV band-pass. 
In pointing mode (on axis) the angular resolution of eROSITA is $18\arcsec$ (HEW) whereas in 
survey mode it is $26\arcsec$ (FOV averaged). Source location accuracy is of the order of 
$4\farcs5$ ($1\sigma$). The second instrument onboard SRG is the Russian X-ray 
concentrator {\sc Mikhail Pavlinsky ART-XC} (Astronomical R\"ontgen Telescope -- X-ray Concentrator) 
\citep{2018SPIE10699E..1YP}, which is sensitive in the hard X-ray band from 4 up to 30 keV, making 
it complementary to the eROSITA soft band.

 \begin{figure}
 \centering
  \includegraphics[width=\columnwidth,clip=true]{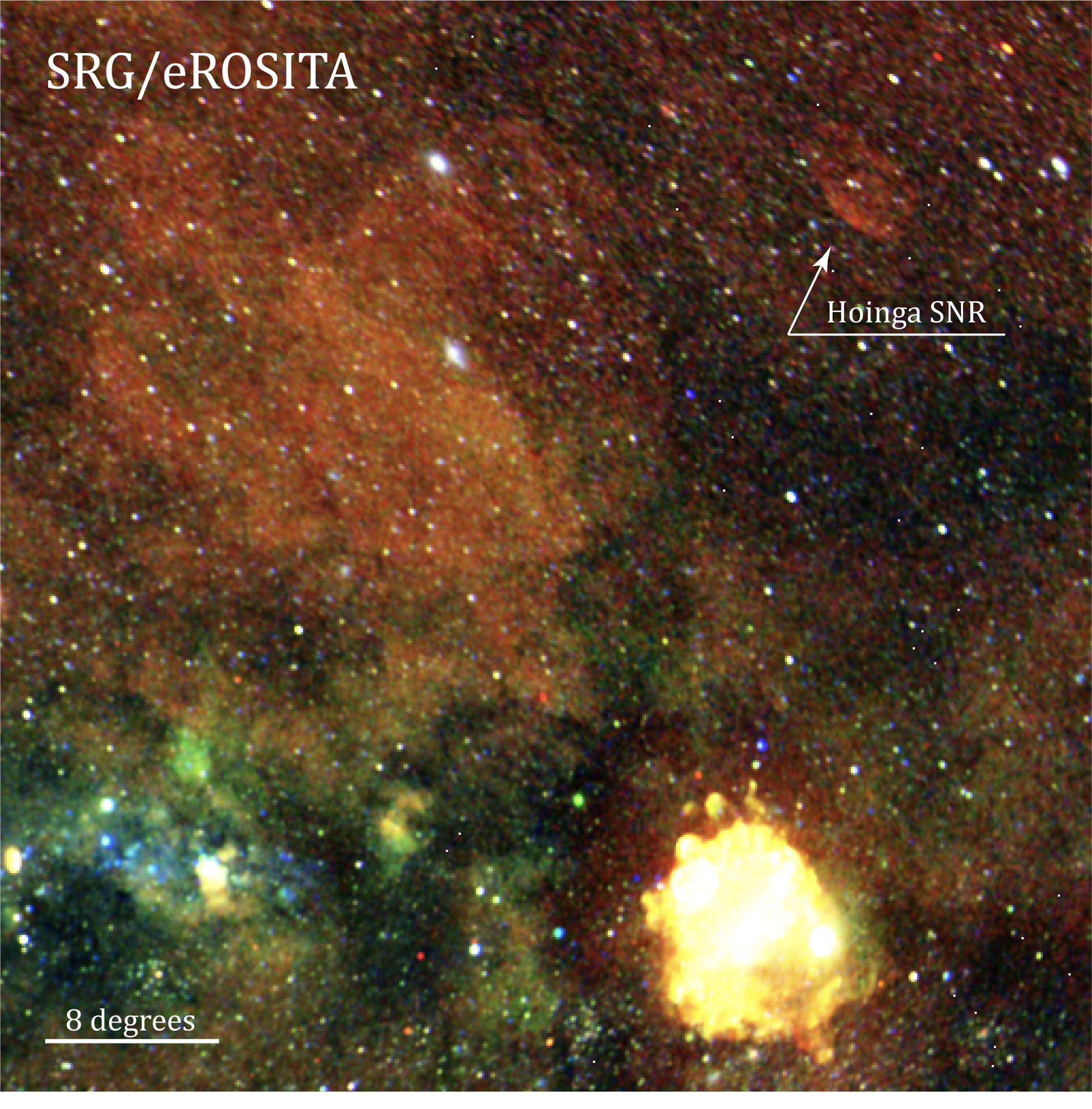}
    \caption{Cutout of the SRG/eROSITA all-sky survey image from eRASS1 data. The image
    shows, among many other sources, the extended X-ray emission from the $24\degr$ 
    diameter large Antlia Loop in its upper left quadrant and the emission 
    from the Vela SNR in its lower right. The emission from the 
    Hoinga SNR in the upper right quadrant of the image is indicated. The image is an 
    Aitoff projection of photons that have been colour-coded according to their 
    energy (red for energies 0.3--0.6 keV, green for 0.6--1 keV, blue for 1--2.3 keV).
    The image was smoothed with a $10\arcmin$ FWHM Gaussian filter and is publicly 
    available on the internet.}
    \label{fig:Fig1}
\end{figure}

The SRG was launched into an L2 orbit on July 13, 2019, with a Russian Proton-M launch vehicle. 
After a three-month calibration and science verification phase it started its first all-sky 
survey on December 13, 2019. With a scan rate of $0.025$ deg/s, a spacecraft revolution 
duration of 4\,h and a central FOV passage time of about 40 s \citep{Predehl2020a}, each 
survey takes 6 months to complete. eROSITA is supposed to take eight all-sky surveys over a
time period of 4 years. 

The X-ray data we report here were taken during the first eROSITA all-sky survey eRASS1, 
completed on June 12, 2020. As the main science driver of the SGR mission is to explore the 
nature of dark energy, its orbit was chosen so that the Ecliptic poles get the deepest exposure, 
leading to an exposure of the Galactic plane which is of the order of $\sim 200-300$ seconds per 
survey.

First results of eRASS1, including a fascinating, detail-rich, three-energy-band colour-coded 
image of the 0.3-8.0 keV X-ray sky, were recently released. The survey represents the deepest 
view of the whole X-ray sky today and led to the discovery of the large-scale symmetric hot-gas 
structures in the Milky Way halo, called `eROSITA Bubbles' \citep{Predehl2020b}, among many other 
exciting results. 

Searching this survey map for unknown extended sources revealed the existence of a new SNR at Galactic 
coordinates $l=249\fdg5$ and $b=24\fdg5$, labelled G249.5+24.5 
which we dub Hoinga\footnote{In honor of the first author's hometown Bad H\"onningen am Rhein: Hoinga
was its medieval name.}. Figure \ref{fig:Fig1} depicts a colour-coded image of the relevant sky region
which shows Hoinga with its neighbours the Antlia Loop and the Vela SNR.

The data we use in our analysis were processed by the eSASS (eROSITA Standard Analysis Software) 
pipeline and have the processing number $\#946$. For the data analysis we used eSASS version 201009 
(released on October 9th, 2020)\footnote{cf.~https://erosita.mpe.mpg.de/}. Within the eSASS pipeline, 
X-ray data of the eRASS sky are divided into 4700 partly overlapping sky tiles of 
$3\fdg6 \times 3\fdg6$ each. These are numbered using six digits, three for RA and three for Dec,
representing the sky tile centre position in degrees. The majority of Hoinga's emission falls into the eRASS1 sky 
tiles numbered 142108, 146108, and 143105 whereas the six surrounding sky tiles (145111, 142111, 139111, 
139108, 146105, 140105) needed to be included for complete coverage of the remnant. Hoinga was 
observed in eRASS1 between 15 and 22 May, 2020, in a total of 29 telescope passages, resulting in an 
unvignetted averaged exposure time of approximately 240 s.    

\begin{figure}
 \centering
  \includegraphics[width=\columnwidth,clip=true]{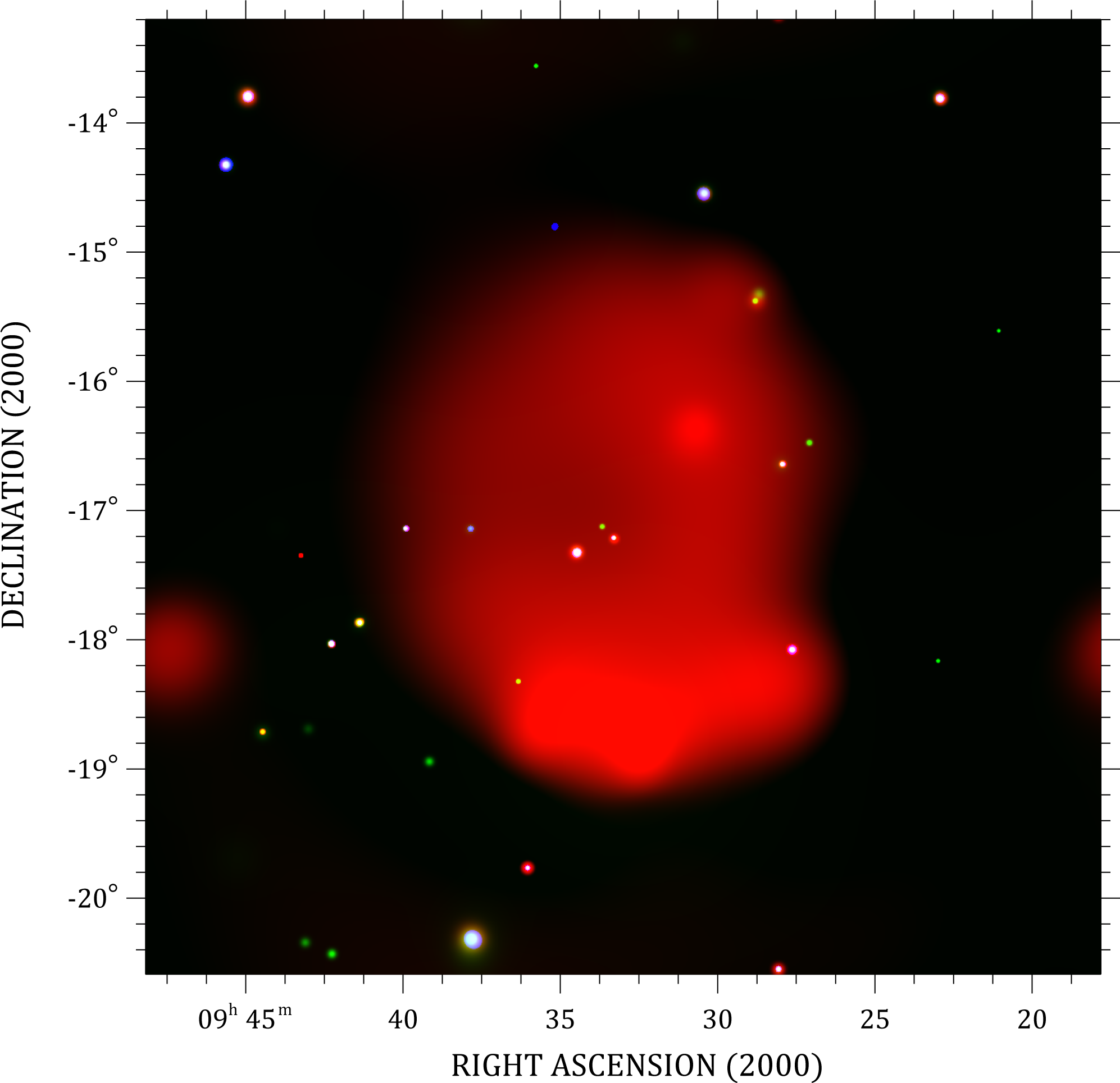}
    \caption{Hoinga SNR as seen in the eROSITA all-sky survey eRASS1. Photons to produce
     this $7\fdg5 \times 7\fdg5$ image were colour coded according to their energy
     (red for energies 0.2--0.7 keV, green for 0.7--1.2 keV, blue for 1.2--2.4 keV). 
     An adaptive kernel smoothing algorithm was applied to the images in each
     energy band.}
    \label{fig:Fig2}
\end{figure}

Figure \ref{fig:Fig2} shows an RGB image of the remnant which has been colour coded according to 
the energy of the detected photons. To produce the image, we first created images for the three
energy bands 0.2--0.7 keV, 0.7--1.2 keV, and 1.2--2.4 keV, respectively. The spatial binning in these 
images was set to $26\arcsec$ which reflects eROSITA's FOV averaged angular resolution 
in survey mode. Data from all seven telescopes were used as we did not notice a significant impact 
of the light leak in TM5 and TM7. In order to enhance the visibility of Hoinga's diffuse emission 
in these images whilst leaving point sources unsmoothed to the greatest possible extent we applied 
the adaptive kernel smoothing algorithm of \cite{2006MNRAS.368...65E} with a Gaussian kernel of
$4.5 \sigma$.   

The image analysis clearly reveals that Hoinga's X-ray emission is very soft. The majority of 
its emission is detected in the 0.2--0.7 keV band, leaving the remnant undetected in eRASS1 
above 0.7 keV. The shape of Hoinga appears largely circular except for the remnant's west side for 
which no emission is detected by eROSITA. The morphological structure of the  remnant is clearly 
centre-filled without a distinct shell-brightening structure. However, its soft X-ray emission 
slightly brightens towards the southern direction with a knot-like structure (cf. also Fig. \ref{fig:Fig3}). 
We will explore this region in more detail when additional eROSITA data become available.
To determine the geometrical centre of the remnant we fitted an annulus to 
the outer boundary of its X-ray emission. In right ascension and declination the remnant 
centre is then found to be at RA = 09:31:53.47, Dec = $-17$:01:36.7 (J2000), which according 
to the eROSITA naming convention assigns it the catalogue name 1eRASS J093153.47--170136.7.

\subsection{Hoinga in the ROSAT All-Sky Survey\label{RASS}}

\begin{figure}
 \centering
  \includegraphics[width=\columnwidth, clip=true]{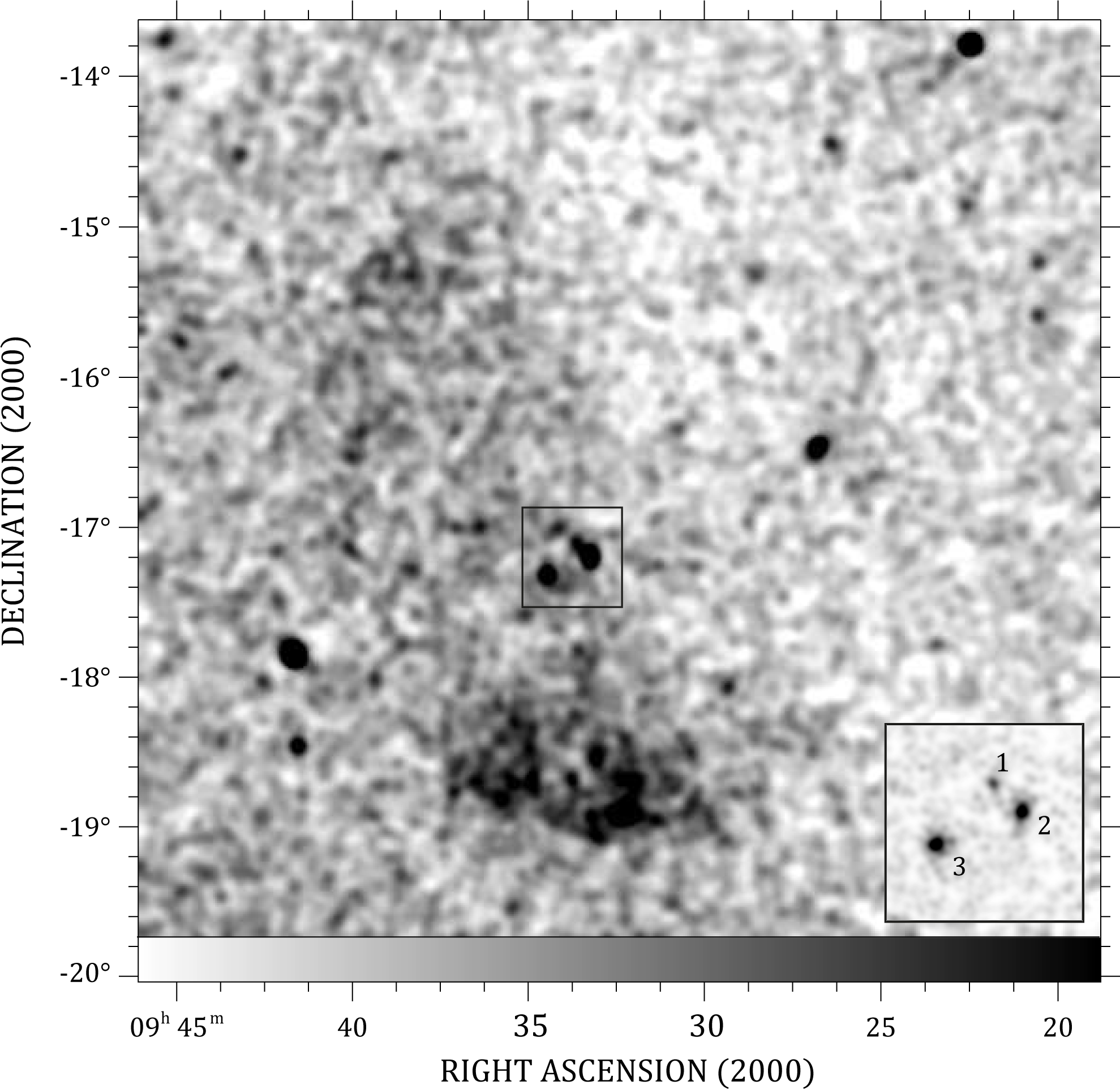}
    \caption{The Hoinga SNR as seen in the ROSAT all-sky survey. Photons to produce
     this image have been selected from within the 0.1--0.7 keV energy band. A Gaussian
     smoothing filter with x,y-$\sigma$ = 3 image pixel was applied in order to enhance
     the visibility of the diffuse emission. The gray
     scale colors are distributed so that white corresponds to a pixel intensity value of
     0.09 and black to 0.45 cts/pixel. The image is vignetting and deadtime corrected though
     no exposure correction was applied. The inset in the lower right corner shows a
     $40\arcmin \times 40\arcmin$ zoom to the region of the X-ray sources located slightly to east
     of the remnant's geometrical centre.}
    \label{fig:Fig3}
\end{figure}

After the discovery of Hoinga in eRASS1 data we went back to the archival ROSAT all-sky survey 
(RASS) to check whether the remnant was detected. The ROSAT RASS was performed between June 1990 
and August 1991, almost exactly 30 years before eRASS1. The ROSAT PSPC (position-sensitive 
proportional counter), which was in the focal plane during the survey, was sensitive in the 0.1-2.4 keV 
energy range \citep{2003NIMPA.515...65P}. The angular resolution in the survey was $45\arcsec$. 
RASS data are divided into 1378 partly overlapping sky tiles, each covering 
$6\fdg4 \times 6\fdg4$ of the sky. Hoinga is located in the RASS data with the
sequence numbers 932025, 932026. It was observed between November 11-18, 1990.  After applying the
standard ROSAT data processing using the Extended Scientific Analysis Software \textsc{EXSAS}
\citep{1991eusg.book.....Z}, we created images from the photons in the 0.1--0.7 keV,
0.7--1.2 keV, and 1.2--2.4 keV energy bands. While there is no emission seen in the medium 
and hard bands, the soft-band image clearly shows a hint of circular shaped soft X-ray emission. 
As in the eROSITA data, its soft X-ray emission is brighter toward 
the south. Figure \ref{fig:Fig3} shows the RASS soft-band image of the relevant sky
region. The effective survey exposure in the image varies from about 480\,s at the eastern side 
of the remnant to about 474\,s near to its central region and 380\,s at its western side. ROSAT's 
scan direction imprint in that sky region is clearly visible in the image by the slightly 
inhomogeneous exposure, from  approximately the southwestern to the northeastern direction.

\subsection{X-ray point sources within the Hoinga SNR}
In order to identify a possible compact remnant associated with Hoinga we applied a source 
detection to the ROSAT and eROSITA survey data. The point sources detected in both
surveys along with their properties are summarised in Table \ref{tab:Tab1}. The eROSITA
68\% position uncertainty for point sources detected in the all-sky survey eRASS1 is 
4\farcs5; for the ROSAT survey it is $13\arcsec$ \citep{1999A&A...349..389V}. 
In Figures \ref{fig:Fig2} and \ref{fig:Fig3}, three point sources can be seen slightly to the 
east of the remnant geometrical centre, though rather centred with respect to the diffuse 
X-ray emission. While the positions of sources $\#$2 and $\#$3 remain unchanged  within 
the errors in ROSAT and eROSITA data, source $\#$1 is found to have an offset towards the 
southeast of almost 20\arcsec. For the purpose of a further source 
identification, we correlated the eROSITA positions with various radio and optical catalogues, 
for example NVSS \citep{1998AJ....115.1693C} and Gaia DR2 \citep{2018A&A...616A...1G}.

For the eROSITA sources $\#2, \#3, \#5, \#6$ we found a convincing positional match to a radio 
counterpart in the NVSS catalogue\footnote{https://www.cv.nrao.edu/nvss}. For all sources, we find a
close overlap with optical sources from the Gaia DR2 catalogue\footnote{https://gea.esac.esa.int/archive}.
From the proper motion and parallax information for the potential counterparts, it seems likely 
that sources $\#1, \#2, \#3, \#5,$ and $\#6$ are of extragalactic nature. In contrast, sources 
$\#4,\#7,\#8,\#9,\#10,$ and $\#11$ appear to be likely of Galactic origin, in agreement with their brighter 
optical appearance.

Assuming the identification of the eROSITA source $\#$1 with an extragalactic optical source is correct, 
% #1 is a QSO  https://simbad.u-strasbg.fr/simbad/simid?Ident=QSOB0931-1655
it seems more likely to us that the computed ROSAT RASS position of this faint X-ray source has 
a larger uncertainty than the $13\arcsec$ found on average (68\% confidence) for ROSAT 
RASS sources \citep{1999A&A...349..389V}. Assuming a real offset for source $\#$1 
 would imply a proper motion of $\sim 20\arcsec / 30$ years, which seems unlikely to 
us as we did not find a nearby bright star as optical counterpart. Indeed, of the 11 X-ray sources 
detected within the Hoinga SNR, none have an optical counterpart fainter than the twentieth magnitude in the 
Gaia G-band. Similarly, in the infrared band where the fainter object is found, H $\simeq 16$  and W2 
($4.6\mu$m) $\simeq 14$ in 2MASS and WISE catalogues, respectively. 
We therefore conclude that all 11 X-ray sources are either foreground or background objects which are not
associated with Hoinga.

\begin{table}
 \centering
 \caption{X-ray sources detected within the Hoinga SNR in eROSITA eRASS1 and ROSAT RASS data. 
 The detection significance of the listed sources is $\ge 5\sigma$. The position uncertainty 
 of eROSITA point sources is 4\farcs5 ($1\sigma$ confidence). <MJD> is the Modified Julian Date 
 of the observation in eRASS1 and RASS, respectively. The numbering for the centrally located 
 sources $\#1-\#3$ is reported in the lower-right inset of Figure \ref{fig:Fig3}.}\label{tab:Tab1}
 \begin{tabular}{c c c c} 
 \hline\hline
Source   &     RA (J2000) &  Dec (J2000)  & Obs. Time \\\hline
         &       h:m:s    &   d:m:s       &   < MJD >  \\ \hline
  \hline\\[-2ex]
 \multicolumn{4}{c}{eROSITA eRASS1}\\
  \hline\\[-2ex]
 1 & 09:33:41.096 & -17:09:18.932 & 58987.40282\\
 2 & 09:33:18.088 & -17:14:41.741 & 58987.31946\\
 3 & 09:34:30.071 & -17:21:21.224 & 58987.65274\\
 4 & 09:37:57.489 & -17:10:14.453 & 58988.23630\\
 5 & 09:27:29.469 & -18:06:20.653 & 58986.48550\\
 6 & 09:27:50.522 & -16:40:01.672 & 58985.98614\\
 7 & 09:36:25.814 & -18:21:05.829 & 58988.40257\\
 8 & 09:28:45.469 & -15:24:10.805 & 58985.73667\\
 9 & 09:28:38.097 & -15:21:08.361 & 58985.73682\\
10 & 09:40:02.199 & -17:09:55.614 & 58988.65310\\
11 & 09:26:58.572 & -16:30:06.584 & 58985.73626\\[0.5ex]
 \hline\\[-2ex] 
  \multicolumn{4}{c}{ROSAT RASS}\\
  \hline\\[-2ex]
  1 & 09:33:41.421  & -17:08:59.602 & 48210.59365\\
  2 & 09:33:18.151  & -17:14:39.386 & 48210.52687\\
  3 & 09:34:30.177  & -17:21:17.739 & 48210.86065\\
  4 & 09:37:57.671  & -17:10:07.991 & 48211.72799\\
  7 & 09:36:26.231  & -18:21:05.791 & 48211.82793\\
 10 & 09:40:02.494  & -17:09:58.469 & 48212.26215\\
 11 & 09:26:58.256  & -16:30:01.930 & 48208.55845\\[0.5ex]
 \hline
 \end{tabular}
 \end{table}
  
\subsection{Spectral analysis\label{spectral_analysis}}
In order to properly correct the source spectrum and energy flux for contributions 
from the instrument- and sky-background, we analysed a sky field of about 
$8\degr \times 8\degr$ centred on the remnant. Hoinga's energy spectrum was 
extracted from the eROSITA eRASS1 data by selecting all events recorded within an 
elliptical region of semi-minor and major axis of $2\fdg0$ and $2\fdg35$, 
respectively. The elliptical region was centred at the position RA = 9:32:57.30, 
Dec = $-$16:51:41.00 and tilted by 14\fdg5. \textsc{SAOImage ds9} \citep{2003ASPC..295..489J} 
was used for the definition of the event-selection regions. The background spectrum was 
extracted from a surrounding elliptical ring for which we chose the axes so that 
it did not include events from the remnant itself. The ring had a difference between 
its inner and outer region of 0\fdg3. Events from unrelated X-ray sources 
located within the source or background regions were excluded from the spectral analysis. 

In total, the extracted spectra included 43910 and 28930 counts from the source and 
background regions, respectively, resulting in about 15000 net events. To model Hoinga's 
X-ray spectrum, we used only events from the telescope units TM1, 2, 3, 4, and 6. Events 
from the two telescopes TM5 and TM7 were excluded from the spectral analysis as both 
units suffer light leaks related to the sun--satellite angle, making their soft-response
calibration quite uncertain at that early stage of the 
mission. Model spectra were simultaneously fitted to Hoinga's source and background 
spectra. We used Xspec 12.10.1f \citep{2003HEAD....7.2210D} and applied the 
C-statistics to the fits in which we modelled the source and background spectra 
independently. Of the fitted model spectra, the APEC spectrum from collisionally ionised 
diffuse gas \citep{2012ApJ...756..128F} and the PSHOCK model (Plane-parallel SHOCK plasma 
model with non-equilibrium ionisation) \citep{2001ApJ...548..820B} were found to provide 
fits of equal goodness and with similar spectral parameters to the observed spectrum. We
used the abundance table and the TBabs absorption model from \cite{2000ApJ...542..914W}. For 
the meaning of the fitted spectral parameters, we refer the reader to the Xspec
manual\footnote{https://heasarc.gsfc.nasa.gov/xanadu/xspec/XspecManual.pdf} and references 
therein. 

\begin{figure} 
 \centering
  \includegraphics[width=\columnwidth, clip=true]{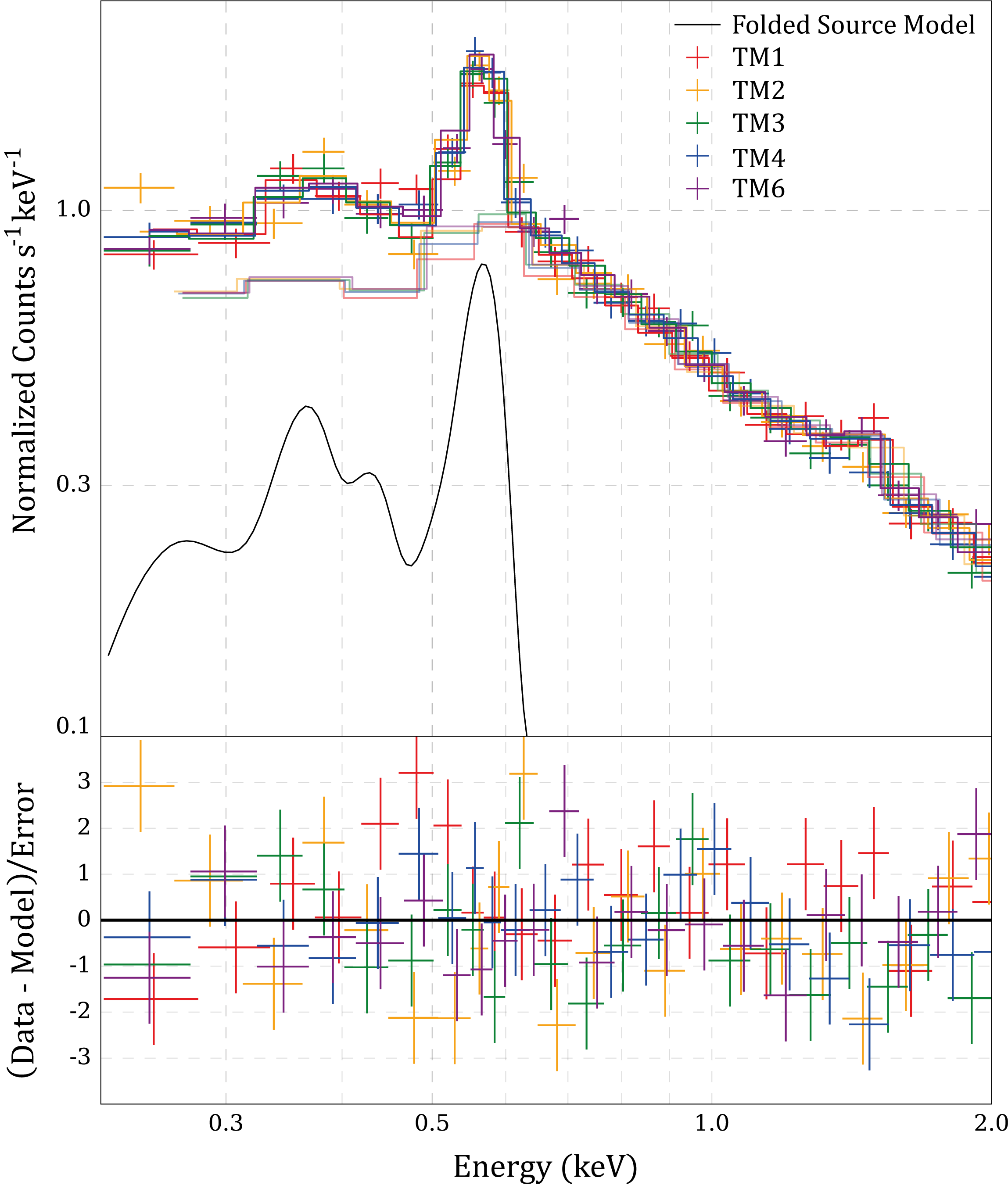}
    \caption{Energy spectrum of the Hoinga SNR as observed with the eROSITA TM1, 2, 3, 4, and 6 
     telescope and detector units and simultaneously fitted to an absorbed APEC spectral
     model ({\it upper panel}). The spectra have been binned for visual clarity and 
     plotting purposes. The signal-to-noise ratio in each bin is $15\sigma$. 
     The folded best-fit APEC spectral model is plotted as a solid black line. Fit 
     residuals are shown in the lower panel. } 
    \label{fig:Fig4}
\end{figure}

\begin{figure} 
 \centering
 \includegraphics[width=\columnwidth, clip=true]{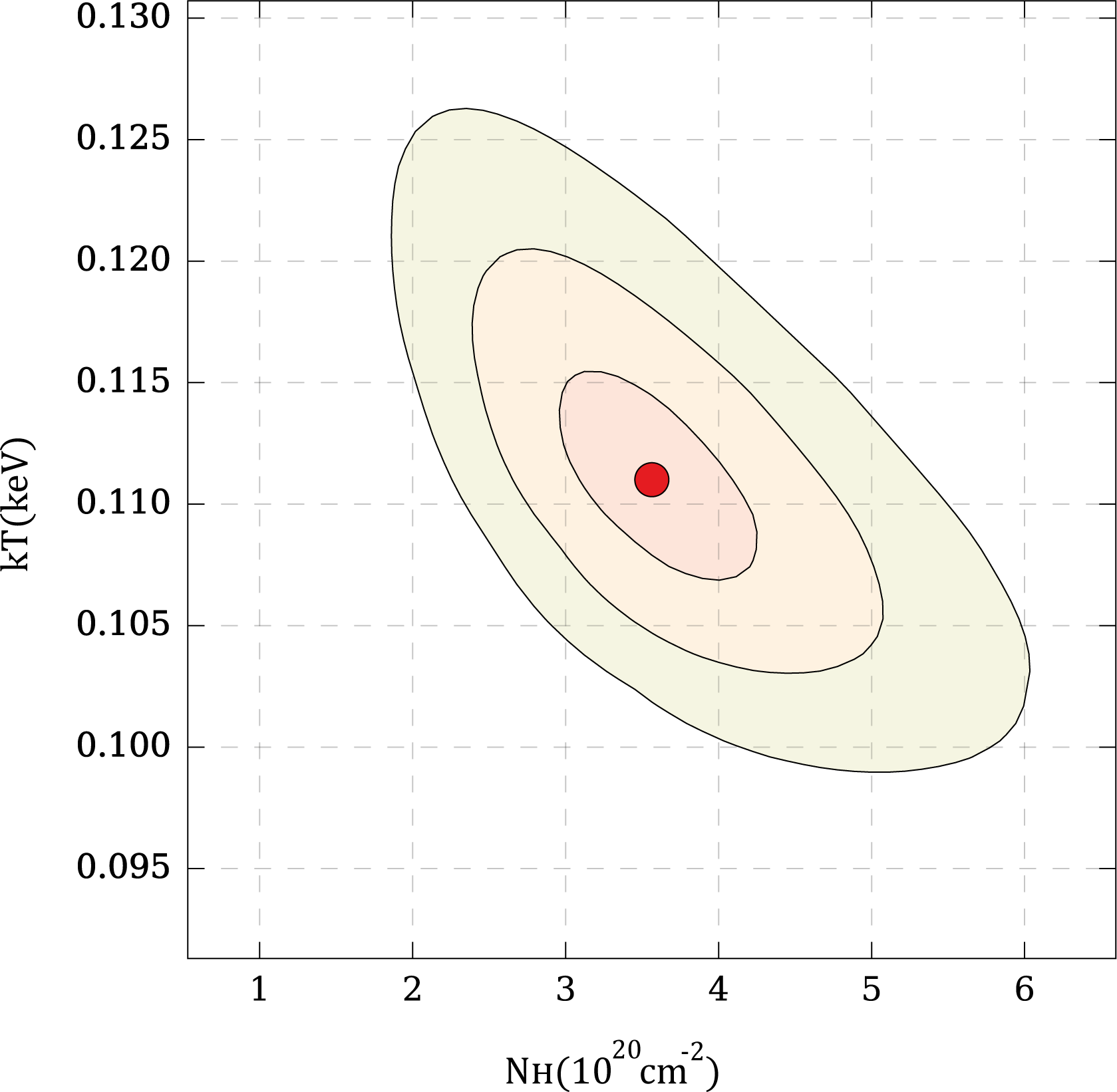}
   \caption{Contour plot showing the relative parameter dependence of the fitted 
   spectral parameters kT (temperature) vs.~N$_H$ (column absorption) for the APEC
   model fit to the energy spectrum of Hoinga. The three contours represent the 
   $1\sigma$, $2\sigma,$ and $3\sigma$ confidence levels for two parameters of interest. 
   The small red dot marks the best-fit position.}  
    \label{fig:Fig5}
\end{figure}

Figure \ref{fig:Fig4} depicts the best-fit APEC model. The model spectrum folded through the 
detector response is shown with a black solid line. Table \ref{tab:Tab2} lists the best-fit 
spectral parameters of both models. Due to the preliminary calibration status of the eROSITA 
instruments at the time of writing, we refrain from giving absolute energy fluxes as obtained from 
the best-fit models. The contour plot shown in Figure \ref{fig:Fig5} gives the parameter 
dependence of the temperature versus the column absorption for the APEC model.

\begin{table}
\centering
 \caption{Best-fit parameters from the fit of APEC and PSHOCK models to the spectrum of Hoinga.
 Errors represent the 68\% confidence range. }\label{tab:Tab2}
\renewcommand{\arraystretch}{1.5}
\begin{tabular}{c c c}
\hline\hline
Parameter & APEC & PSHOCK \\
\hline
$N_{\rm{H}} \, (10^{20} \rm{cm}^{-2})$  & $3.6^{+0.7}_{-0.6}$       & $3.6^{+0.6}_{-1.0}$\\
$kT \, (\rm{keV})$                      & $0.111^{+0.004}_{-0.004}$  & $0.108^{+0.012}_{-0.008}$\\
$\tau \, (10^{11} \rm{s\,cm}^{-3})$     & ...                        & $> 1.1$\tablefootmark{a}\\ 
Normalization                           & $0.17^{+0.03}_{-0.03} $   & $0.13^{+0.06}_{-0.06}$\\ 
C Statistic / d.o.f.                    & $7625.1 / 7347 $           & $7625.1 / 7346 $\\ 
\hline
\end{tabular}
\tablefoot{%\\
\tablefoottext{a}{The ionisation timescale $\tau$ is only weakly constrained by the fitted spectrum, 
which is why we only give a $95\%$ lower limit.}}
\end{table}

\section{Radio observations and data analysis\label{sec:radio}}

\subsection{The Murchison Widefield Array} \label{sec:mwa}

\begin{figure*} 
    \includegraphics[width=18cm, clip=true]{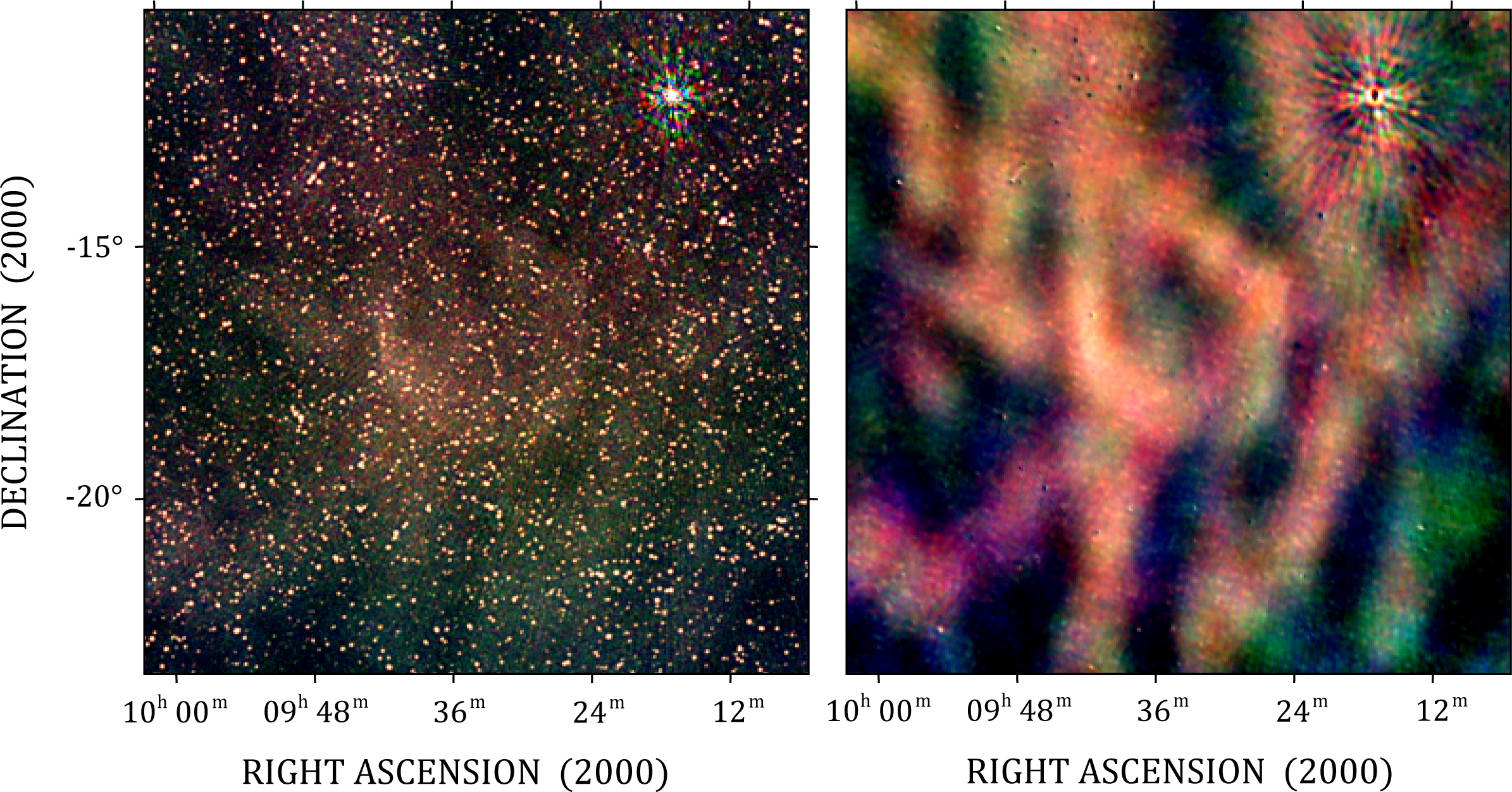}
    \caption{$\sim 10\degr \times 10\degr$ of the region surrounding Hoinga as seen by GLEAM at 103-134\,MHz (R), 
    139-170\,MHz (G), and 170-200\,MHz (B). The left panel shows the image from the data release 
    of \cite{2017MNRAS.464.1146H}, and the right panel shows the region after reprocessing to subtract 
    sources and highlight large-scale structure (see \Sect~\ref{sec:mwa}). Hoinga is visible as an 
    ellipse in the centre of the image; steep-spectrum Galactic cirrus becomes a strong contaminant 
    at these low frequencies and is visible as large-scale filaments around the remnant. The bright 
    source in the northwest is Hydra~A.}
    \label{fig:mwa}
\end{figure*}

The Murchison Widefield Array \citep[MWA; ][]{2013PASA...30....7T,2018PASA...35...33W} is a low-frequency 
radio telescope operating in Western Australia, and is a precursor to the 
low-frequency component of the Square Kilometre Array. The GaLactic and Extragalactic All-sky 
MWA \citep[GLEAM; ][]{2015PASA...32...25W} survey observed the whole sky south of declination 
(Dec) $+30\degr$ from 2013 to 2015 between 72 and 231\,MHz. A major data release covering 24\,402 
square degrees of extragalactic sky was published by \cite{2017MNRAS.464.1146H}, while individual 
studies have published smaller regions such as the Magellanic Clouds \citep{2018MNRAS.480.2743F} 
and parts of the Galactic plane \citep{2019PASA...36...47H}. An important feature of this radio 
survey is its sensitivity to large-scale ($1\degr - 15\degr$) features, which has enabled studies of 
SNRs and \textsc{Hii} regions across a wide range of sizes and the full range of frequencies, 
independent of resolution biases \citep[see e.g.][]{2016PASA...33...20H,2018MNRAS.479.4041S,2019PASA...36...45H}.

Hoinga is visible in the public GLEAM  images\footnote{\href{http://gleam-vo.icrar.org/gleam\_postage/q/form}{http://gleam-vo.icrar.org/gleam\_postage/q/form}} 
but is contaminated by the presence of hundreds of radio sources, the majority of which are likely 
unrelated radio galaxies (left panel of \Fig~\ref{fig:mwa}). To accurately measure the radio flux 
density of Hoinga, we reprocessed 13 two-minute observations spanning 103-231\,MHz from a drift scan 
centred at Dec~$-13\degr$ taken on 2014-03-04, with three or four observations in each 30.72-MHz band, yielding 
integration times of $\approx10$\,minutes per band.
For each observation, we performed the following steps, in each case attenuating the brightness 
of modelled sources using the MWA primary beam model of \cite{2017PASA...34...62S}:

\begin{itemize}
    \item Download the data from the All-Sky Virtual Observatory\footnote{\href{https://asvo.mwatelescope.org/}{https://asvo.mwatelescope.org/}}
    in standard measurement set format, averaged to 40\,kHz and 2\,s frequency and time resolution; 
    \item calculate a first-pass amplitude and phase calibration for each antenna using a sky model 
    comprised of the bright nearby source Hydra~A and the GLEAM catalogue, via the software \textsc{calibrate}, 
    an implementation of the \textsc{MitchCal} algorithm \citep{2016MNRAS.458.1057O};
    \item apply the derived calibration solutions;
    \item use the \textsc{peel} software to remove Hydra~A from the visibilities, with a solution interval of 4\,s;
    \item directly subtract the GLEAM sources from the visibilities using \textsc{subtrmodel};
    \item use the widefield radio imaging package \textsc{WSClean} \citep{2014MNRAS.444..606O} to 
    image the data using natural weighting and multi-scale multi-frequency synthesis over the full 
    30.72-MHz band down to a threshold of three times the local image noise, and then clean the data down to the local image noise in regions found to contain brightness.
\end{itemize}

The ionosphere was found to be in a relatively quiescent state, with minor ($\approx$arcsec) position 
shifts imparted to the radio sources; the images were corrected using \textsc{fits\_warp} 
\citep{2018A&C....25...94H}. For each 30.72-MHz band, the primary-beam-corrected images were 
then mosaicked using \textsc{swarp} \citep{2002ASPC..281..228B}. The resulting image is shown 
in the right panel of \Fig~\ref{fig:mwa}. Hoinga is visible as a pair of arcs of width $\approx1\degr$, 
$5\degr$ apart from one another. The local diffuse Galactic synchrotron is also visible 
as a fainter series of filaments with a similar colour (i.e. spectral index). 

We used the software \textsc{poly\_flux} \citep{2019PASA...36...48H} to measure the total flux 
densities of Hoinga in each band, estimating and subtracting a mean background level. As the 
selection of the boundaries of the SNR is somewhat subjective, we used the tool ten times and 
recorded the average result. The results are shown in \Tab~\ref{tab:radio_results}. The uncertainties 
are estimated at 20\,\%, dominated by the difficulty in selecting the true bounds of the SNR and calculating the true background level of the Galactic cirrus.

\subsection{Haslam}\label{sec:haslam}

\begin{figure*} 
 \includegraphics[width=18cm, clip=true]{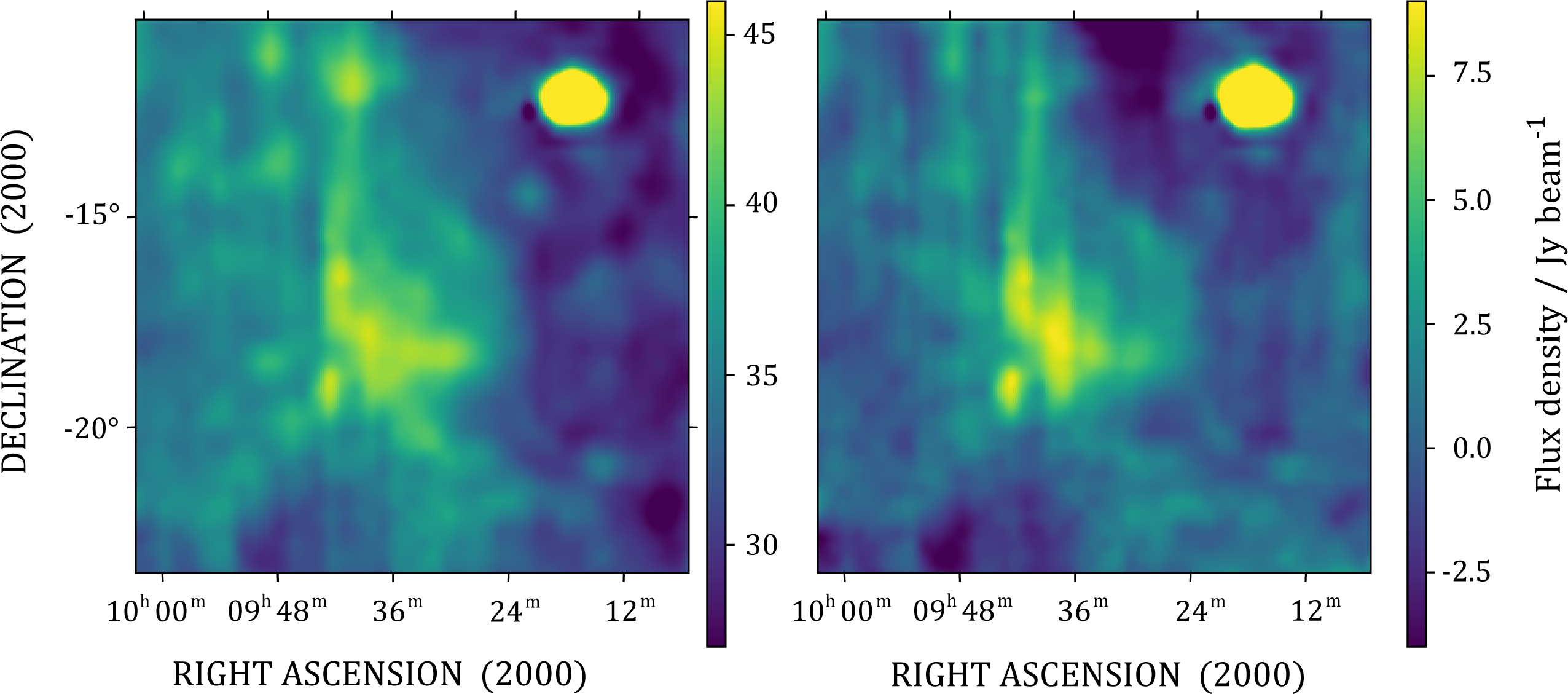}
    \caption{100\,deg$^2$ of the region surrounding the Hoinga SNR as seen 
    at 408\,MHz by the survey by \cite{1982A&AS...47....1H}, after conversion from K to 
    Jy\,beam$^{-1}$. The left panel shows the original image, and the right panel shows 
    the image after source-subtraction and backgrounding, discussed in \Sect~\ref{sec:haslam}. 
    Hoinga is visible as an ellipse in the centre of the image, while Galactic cirrus and 
    scan line artefacts from the Parkes observing strategy dominate the surroundings. 
    The bright source in the northwest is Hydra~A, and subtraction of this source has 
    not been performed.}
    \label{fig:haslam}
\end{figure*}

The all-sky 408-MHz `Haslam' survey was performed with the Green Bank and Parkes Radio 
telescopes and remains the lowest-frequency total-power measurement of the full sky \citep{1982A&AS...47....1H}. 
The Hoinga SNR is visible in the Haslam images (\Fig~\ref{fig:haslam}) but 
the scanning pattern of Parkes is visible as a series of 
vertical lines of varying brightness throughout the image. As this is a total 
power measurement, the largest scale Galactic cirrus features are much brighter than 
Hoinga, leading to a large increase in brightness between the east and west parts of the 
image. The images are also invisibly contaminated by the same radio sources resolved 
in the GLEAM data (\Sect~\ref{sec:mwa}). To mitigate these issues, we used the following 
steps:
\begin{itemize}
    \item model and subtract the GLEAM extragalactic catalogue for this region, 
    extrapolating the source spectra to 408\,MHz, either via their spectral index 
    $\alpha$ as measured by GLEAM or for the fainter sources, by an assumed 
    value of $-0.75$;
    \item determine the average brightness profile over the lower portion of the 
    image (south of Hoinga) as a function of right ascension, and subtract this 
    profile from the full image.
\end{itemize}

This resulted in the right-hand panel of \Fig~\ref{fig:haslam}, where the artefacts 
and contaminating sources have largely been removed. Similarly to the GLEAM data, we 
ran \textsc{poly\_flux} and found that the uncertainty on the final results was dominated 
by the difficulty in subtracting the background, which still has large scan artefacts. 
We therefore conservatively estimate the error at 20\,\%.

We also attempted to use the `de-striped' `de-sourced' version of the Haslam image 
produced by \cite{2015MNRAS.451.4311R}, but Hoinga was invisible in this version, 
possibly because it has similar angular scale to the scanning artefacts, and so was 
removed by the clean-up algorithms employed.

\subsection{CHIPASS}\label{sec:chipass}

\begin{figure*}
   \includegraphics[width=18cm, clip=true]{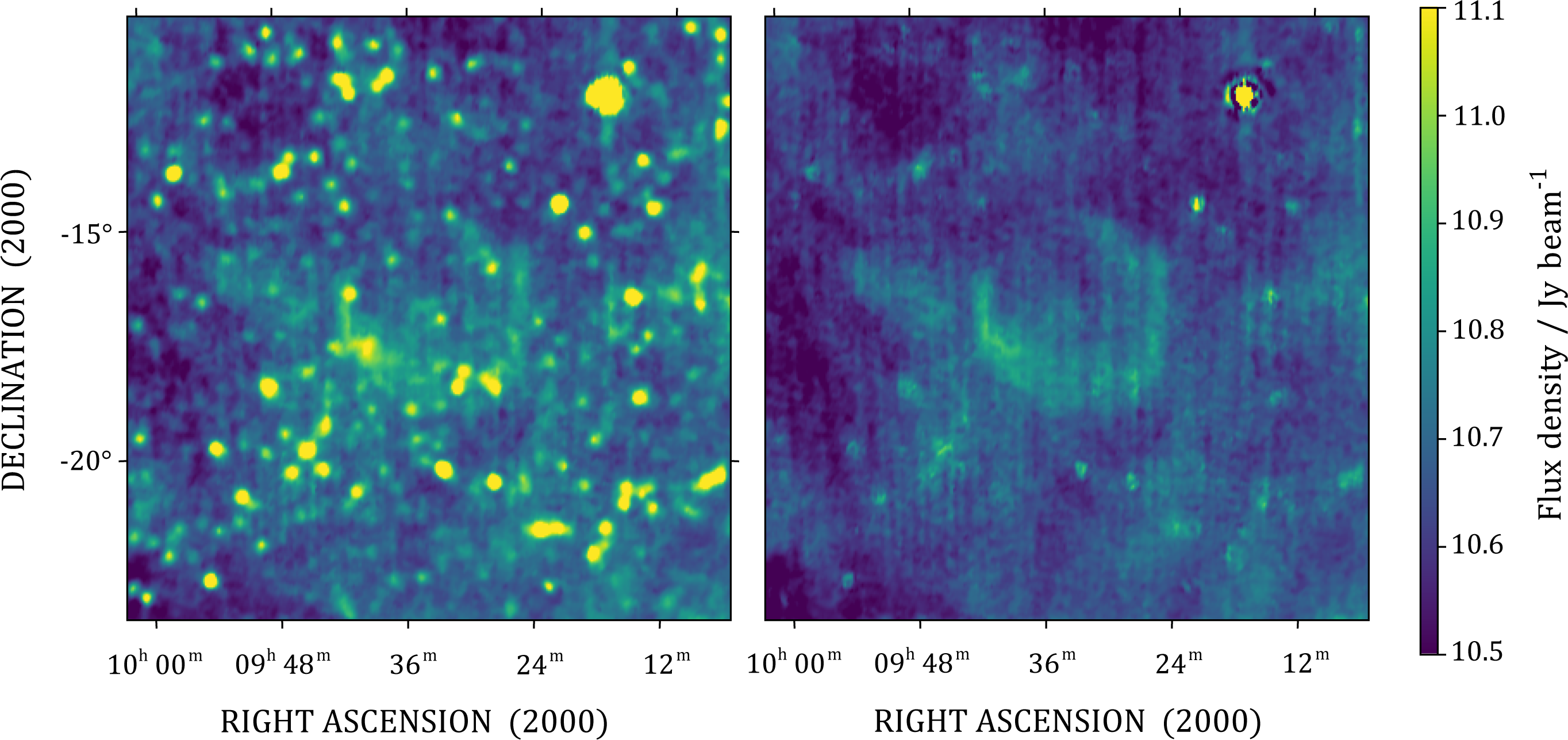}
    \caption{100\,deg$^2$ of the region surrounding Hoinga as seen at 1.4\,GHz by CHIPASS, 
    after conversion from K to Jy\,beam$^{-1}$. The left panel shows the original image, 
    and the right panel shows the image after source subtraction, discussed in 
    \Sect~\ref{sec:chipass}. Hoinga is clearly visible as a crescent-moon in 
    the centre of the image, while Galactic cirrus and residuals around poorly subtracted 
    diffuse sources are visible in the surroundings. The bright source in the northwest 
    is Hydra~A. Faint scan lines are visible from the Parkes observing strategy.}
    \label{fig:chipass}
\end{figure*}

The continuum map of the HI Parkes All-Sky Survey \citep[CHIPASS; ][]{2014PASA...31....7C} 
maps the radio sky at 1.4\,GHz south of Dec~$+25\degr$. We downloaded the data\footnote{\href{https://www.atnf.csiro.au/people/mcalabre/CHIPASS/index.html}{https://www.atnf.csiro.au/people/mcalabre/CHIPASS/index.html}}, and cropped and regridded it to match the 
MWA mosaics (left panel of \Fig~\ref{fig:chipass}). We selected sources within $15\degr$ 
of Hoinga from the NRAO VLA Sky Survey \citep[NVSS; ][]{1998AJ....115.1693C}, convolved them 
to match the CHIPASS resolution, and produced an output FITS image in the same sky frame as 
the regridded CHIPASS data. We subtracted the NVSS model from the CHIPASS image, producing 
the right panel of \Fig~\ref{fig:chipass}. We used \textsc{poly\_flux} to measure the flux 
density of Hoinga, shown in \Tab~\ref{tab:radio_results}. The errors are dominated by the 
selection of the region for subtraction, and after repeated measurements, we estimate this 
at about 5\,\%, which is 1\,Jy.

\subsection{S-PASS}\label{sec:spass}

The S-Band Polarization All Sky Survey \citep[SPASS ;][]{2019MNRAS.489.2330C} is a survey 
of polarized radio emission over the southern sky at Dec $<-1\degr$ using the Parkes 
radio telescope at 2.3\,GHz. Unlike for CHIPASS (\Sect~\ref{sec:chipass}) there is no 
independent catalogue of extragalactic radio sources at 2.3\,GHz. \cite{2017PASA...34...13M} 
derived a catalogue of radio sources from a version of the S-PASS images where the large-scale 
emission had been filtered out, with slightly worse resolution (10\farcm75) than the published 
images (8\farcm9). This catalogue is not as sensitive as and is more confused than NVSS yielding 
a source density equal to 3\,\% that of NVSS.

\begin{figure*} 
   \includegraphics[width=18cm, clip=true]{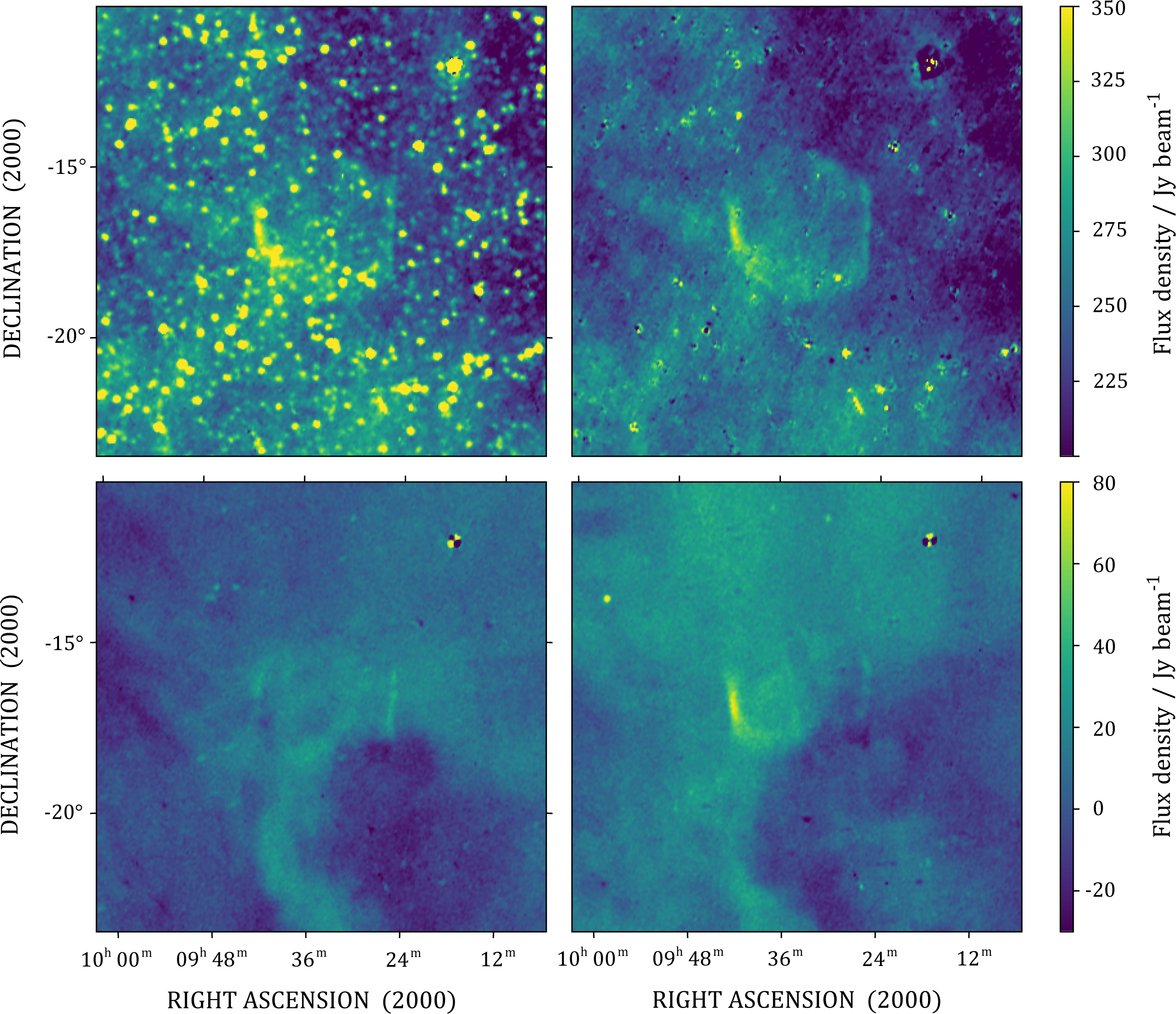}
    \caption{100\,deg$^2$ of the region surrounding Hoinga as seen at 2.3\,GHz by SPASS. 
    The top-left panel shows the Stokes~I image, and the top-right panel shows the 
    image after source subtraction, discussed in \Sect~\ref{sec:spass}. Hoinga is clearly 
    visible as a filled ellipse in the centre of the image, while Galactic cirrus and 
    residuals around poorly subtracted diffuse sources are visible in the surroundings. 
    The bottom left and bottom right panels show the Stokes~Q and U images, respectively.}
    \label{fig:spass}
\end{figure*}

We therefore use NVSS to create the local model of sources to subtract. To obtain spectral
indices for each source, we use the catalogue produced by \cite{2018MNRAS.474.5008D}; 
for sources without a listed spectral index, we use the median local value of $\alpha=-0.75$. 
Subtracting this model from the S-PASS data results in the right-hand panel of \Fig~\ref{fig:spass}. 
Running \textsc{poly\_flux} repeatedly we find more consistent results than for CHIPASS; 
the uncertainty is most likely dominated by the less clean source subtraction. The residual 
RMS after source subtraction in a given beam is $\approx20$\,mJy\,beam$^{-1}$; Hoinga 
subtends 256 SPASS beams; the error is therefore estimated as 0.32\,Jy.

As S-PASS is a polarisation survey, we can also examine the Stokes~Q and U images 
of the region, which indicate the degree of linear polarisation at angles of $\pm90\degr$
and $\pm45\degr$, respectively. Figure~\ref{fig:spass} shows that the brightest parts of
the shell (left and right `limbs') show clear linear polarisation, which is what would 
be expected from a middle-aged SNR shell with a large shock compression ratio. These also 
correspond to flatter parts of the SNR shell, perhaps indicating a local increase in gas density.

\begin{table}
    \centering
    \caption{Integrated flux densities of Hoinga measured from the radio data described in \Sect~\ref{sec:radio}. 
    Measurements were made on images where contaminating sources and background had been 
    removed using the software \textsc{poly\_flux}.}
    \label{tab:radio_results}
    \begin{tabular}{c|ccc}
        \hline
      Survey   &  Frequency / & Resolution / &  Flux density / \\
             &  MHz  & \arcmin & Jy \\
    \hline
       GLEAM  & 118 & $7.8\times6.6$ & $115\pm23$ \\
       GLEAM  & 154 & $6.0\times5.0$ & $100\pm20$ \\
       GLEAM & 185 & $5.0\times4.2$ & $90\pm18$ \\
       GLEAM & 215 & $4.2\times3.6$ & $80\pm16$ \\
       Haslam & 408 & 51 & $60\pm10$ \\
       CHIPASS & 1400 & 14.4 & $19.7\pm1.0$ \\
       SPASS   & 2300 & 8.9 & $15.2\pm0.3$ \\
       \hline
    \end{tabular}
\end{table}

\section{Constraints on $^{44}\rm Ti$ emission from INTEGRAL}

Explosive nucleosynthesis in SNe is considered the main driver of Galactic, chemical evolution.
Its imprints can be readily investigated by observing the $\gamma$-rays emitted in the
decay from freshly synthesized, radioactive nuclei. With a half-life of 58.9 years,
the abundantly produced $^{44}$Ti is an ideal candidate with which to study nucleosynthesis
imprinting in young SNRs.

In core collapse supernovae (ccSN) $^{44}$Ti is mainly produced during the $\alpha$-rich
freeze-out \citep{Woosley1973} deep in the central region, where the nucleosynthesis yields
are strongly dependent on the thermodynamic conditions \citep{Magkotsios2010,Hermansen2020}.
While models of ccSN fail to robustly produce explosions in a wide stellar mass range so
far, it appears safe to assume that asymmetries are required to drive successful explosions.
Depending on the applied, simplified explosion scheme, the predicted $^{44}$Ti ejecta yield
can vary in the range $10^{-5} - 10^{-4}$\,M$_\odot$, depending also on the initial mass of the
exploding star \citep{Timmes1996,Wanajo2018,Limongi2018}.

In contrast, thermonuclear SNe (type Ia) show a larger diversity in the predicted
$^{44}$Ti ejecta masses. Multiple scenarios leading to the disruption of a white dwarf
star are considered viable, as the progenitors of these explosions have not yet been
unambiguously identified. For the standard model, involving a centrally ignited 
Chandrasekhar-mass white dwarf star, $^{44}$Ti ejecta masses range between $10^{-6}$ and $10^{-5}$\,M$_\odot$
\citep{Maeda2010,Seitenzahl2013,Fink2014}. However, in the double-detonation scenario,
ejecta masses of $10^{-3} - 10^{-2}$\,M$_\odot$ are possible \citep{Fink2010,Woosley2011,Moll2013},
where some exotic models even predict $^{44}$Ti masses of up to 0.1\,M$_\odot$ \citep{Perets2010,Waldman2011}.

Evidence for the production of $^{44}$Ti can be obtained by measuring the decay radiation in
the decay chain of $^{44}\rm Ti$ $\rightarrow$ $^{44}\rm Sc$ $\rightarrow$ $^{44}\rm Ca$. The
dominant decay lines are emitted at 68 and 78\,keV during the $^{44}$Ti decay with a half life
of 58.9\,years \citep{Ahmad2006} and at 1157\,keV in the subsequent $^{44}$Sc decay with a half life
of 4\,hours \citep{Audi2003}. Photons are emitted with a probability (branching ratio) of 93.0\,$\%$,
96.4\,$\%,$ and 99.9\,$\%$ per decay, respectively \citep{Chen2011}

Here, the spectrometer SPI \citep{Vedrenne2003} on INTEGRAL \citep{Winkler2003} is used to 
search for the decay radiation in both subsequent decay steps in the Hoinga SNR. We use the 
\textit{spimodfit} analysis tool \citep{Strong2005,Halloin2009} to extract the spectrum in the 
relevant energy ranges 50--100~keV and 1100--1200~keV from the raw SPI data. The spectrum is 
extracted assuming an extended source of emission modelled by a circular region of 2\fdg2 radius
with a constant surface brightness. A detailed description of SPI analysis and robust background 
modelling can be found in \citet{Diehl2018,Siegert2019,Weinberger2020}.\\

The extracted spectrum is modelled with a general continuum and a variable number of Gaussian-shaped 
decay lines given by
\begin{equation}
        LS(E;E_0,F_0,\sigma) = \frac{F_0}{\sqrt{2\pi}\sigma}\cdot\mathrm{exp}\left(\frac{(E-E_0)^2}{2\sigma^2}\right) + A_0\cdot \left(\frac{E}{E_C}\right)^{\alpha}
,\end{equation}

where $F_0$ is the measured line flux, $E_0$ is the energy of the Doppler-shifted line centroid, 
and $\sigma$ is the line width. As we expect a low signal-to-noise ratio for the decay lines, we 
search for a combined signal in all lines simultaneously, that is we assume that the branching ratio 
corrected fluxes, Doppler shifts, and broadening are identical in all lines. Due to the presence 
of a complex of strong background lines between 50 and 65\,keV induced by germanium, we excluded 
the 68\,keV line in the analysis.

We find no significant flux excess in the vicinity of the 78 or 1157\,keV line or in the combined 
line analysis. As the broadening of the 78 or 1157\,keV lines is related to the expansion velocity 
of the $^{44}$Ti-containing ejecta and determines the size of the selected background region, we 
deduce a $3\sigma$ upper flux limit of $9.2 \times 10^{-5}$\,ph\,cm$^{-2}$\,s$^{-1}$ by assuming 
an expansion velocity of 4000\,km\,s$^{-1}$ \citep{1998ApJ...495..413N,2015A&A...574A..72D}. This 
expansion velocity translates into a line broadening of $\approx 2$\,keV FWHM at 78\,keV and 
$\approx 20$\,keV FWHM at 1157\,keV, respectively. 

\section{Summary and Discussion \label{Sum}}
Using data from the first  SRG/eROSITA observatory all-sky survey we discovered 
one of the largest SNRs in the sky. Despite 95\,\% of SNR discoveries being 
made at radio wavelengths, and its clear existence in multiple radio surveys, we conclude 
that Hoinga was missed by previous searches for several reasons. 
Firstly, its location at high Galactic latitudes; most radio searches have focused
on low latitudes, where the density of SNRs is expected to be highest. Another reason for 
not noticing it in previous X-ray and radio surveys is its total flux density. Although it is 
large, its surface brightness is relatively low. As it has very little fine-scale structure, 
it also does not appear at all in most interferometric maps. In single-dish radio images, it 
is visibly contaminated by about $100$ extragalactic radio sources, with many more below 
the sensitivity and confusion limits, meaning that its diffuse radio emission remained uncovered. 
Hoinga is nearly the largest SNR ever detected at radio wavelengths, subtending 
$\approx275\arcmin \times 265\arcmin$, and comparable in size to the largest detected object, 
G\,65.3+5.7: it was therefore outside the bounds of what was expected and was therefore not visually detected. 
Finally, its similar angular scale and structure to the diffuse Galactic synchrotron 
makes it less obvious than smaller and brighter sources.

\begin{figure}
    \centering
   \includegraphics[width=\columnwidth]{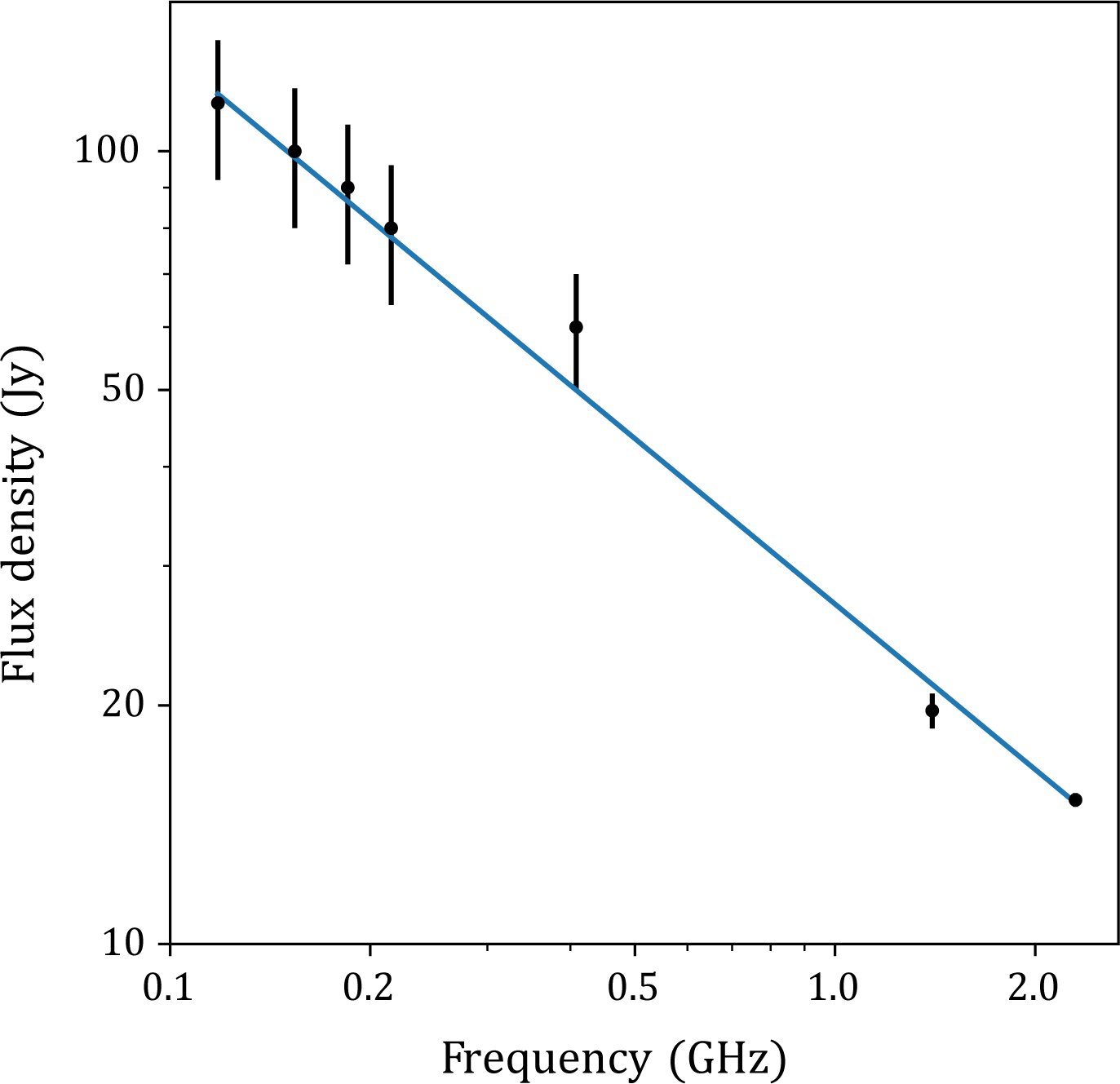}
    \caption{Radio SED of the total flux density of Hoinga as measured by the
    surveys discussed in \Sect~\ref{sec:radio}. Black points show the data
    from \Tab~\ref{tab:radio_results}; the blue line shows a
    least-squares weighted fit to the data, yielding $S_\mathrm{1GHz}=26.85\pm0.03$\,Jy
    and $\alpha=-0.69\pm0.08$ for $S_\nu\propto\nu^\alpha$.}
    \label{fig:SED}
\end{figure}

The clear shell structure, particularly evident in \Figs~\ref{fig:Fig1} and~\ref{fig:chipass},
indicates it is likely to be a classic shell-type SNR that is not centrally powered, and its 
highly circular nature indicates that it is expanding into a region of relatively uniform
density. Figure~\ref{fig:SED} shows the radio flux densities plotted as a function of frequency, 
with a fitted spectral index of $\alpha=-0.69\pm0.08$, for $S_\nu\propto\nu^\alpha$. This 
radio spectral energy distribution indicates that non-thermal synchrotron emission dominates 
the radio spectrum, again consistent with a shell-type SNR.

A distance to the SNR would enable transformation of our measurements into physical 
properties. \cite{2015A&ARv..23....3D} discuss the challenge of estimating the distance 
of radio-detected SNRs; a method that does not rely on additional observations is to search
for nearby neutron stars that appear as pulsars and may have formed at the same time as 
the SNR, and using their dispersion measure in combination with electron density models of
the Galaxy to determine their distance.

We used the Australia Telescope National Facility pulsar catalogue v1.59 
\citep{2005AJ....129.1993M}\footnote{\href{http://www.atnf.csiro.au/research/pulsar/psrcat}{atnf.csiro.au/research/pulsar/psrcat/}} 
to search for known radio pulsars within $20\degr$ of Hoinga's geometrical centre, 
but found none with attributes that would indicate a clear association. From the group 
of pulsars located in the region of interest we excluded possible matches on the basis of:
\begin{itemize}
    \item period $P < 10$\,ms, indicating a recycled origin;
    \item characteristic ages ($\frac{P}{2\dot{P}}$) $>45$\,Myr, which would be extremely 
    inconsistent with a SNR age of $<0.5$\,Myr;
    \item measured proper motion inconsistent with having a common centre of origin;
    \item measured dispersion measure inconsistent with a nearby location.
\end{itemize}
This avenue is therefore unpromising, but because the coverage of pulsar surveys is denser 
at low Galactic latitude, a pulsar could have been missed by existing observations, and follow-up 
observations within the SNR shell may yet reveal a counterpart. Assuming a distance of $\sim 500$ 
pc, a remnant NS with a transversal speed of the order of 1000 km s$^{-1}$ would have by now reached 
the SN shell if the explosion happened $\sim 17000$ years ago. This speed is not unrealistic, 
albeit at the far side of the velocity distribution (see e.g. \cite{1993Natur.362..133C, 2005ApJ...630L..61C, Becker09}). 
We will investigate this possibility in future work.

In the absence of a measured distance, we can use the morphological and brightness 
properties of the SNR to infer limits on the physical characteristics. Studies of the
Magellanic Clouds and other Local Group galaxies show that SNR 1.4-GHz luminosities 
typically have values in the range $5\times10^{14} < L_\mathrm{1.4GHz} < 10^{17}$\,W\,Hz$^{-1}$ 
\citep[e.g.][]{1998ApJ...504..761C}. Assuming that Hoinga is 
more luminous than $5\times10^{14}$\,W\,Hz$^{-1}$, we can obtain a limit on its distance 
from Earth by $\sqrt{\frac{L_\mathrm{1.4GHz}}{4\pi S_\mathrm{1.4GHz}}}$, i.e. $D>450$\,pc. 
Additionally, radio SNRs do not typically have diameters greater than 100\,pc 
\citep[][]{2010MNRAS.407.1301B}. If we assume that Hoinga has a diameter $<100$\,pc, 
by geometry its distance from Earth must be $D<1.2$\,kpc.
This also gives rise to a  luminosity limit of $L_\mathrm{1.4GHz}<1.3\times10^{16}$\,W\,Hz$^{-1}$, 
which puts Hoinga on the lower end of the SNR luminosity distribution. We note that other 
high-latitude SNRs have also been found to have unusually low brightness compared to those 
at low latitudes; see e.g. G181.1$+9.5$ \citep{2017A&A...597A.116K} and G\,0.1$-9.7$
\citep{2019PASA...36...45H}.
% A note about \Sigma D.....
%L = S x 4pi x D^2
%S = 20 x 10^-26 W/Hz/m^2
%D < 1.2 x 3.1 x 10^19 m
%L <= 1.3 x 10^16 W/Hz
% Not a strong constraint, but nice and consistent with what you'd expect
%
% If we assume it has to be more luminous than 5x10^14 W/Hz
% D > sqrt(L/S4pi)
% D > 450 pc

If we compare the remnant with other nearby SNRs such as the Vela SNR, which is also known to 
have an extent of 8\fdg8 and a thermal X-ray spectrum with gas temperatures in the range of 
0.2--0.7~keV, a simple scaling law puts Hoinga at twice the distance of the Vela SNR, which 
is about 500\,pc. The column absorption through the Galaxy into the direction of Hoinga is 
$6 \times 10^{20}$\,cm$^{-2}$ \citep{1990ARA&A..28..215D}. The values found from our X-ray 
spectral fits are of the order of $N_{\rm{H}}=3.6^{+0.7}_{-0.6} \times 10^{20}$\,cm$^{-2}$ 
which gives another indication for Hoinga being a nearby SNR.

If we assume that the column density derived in Section \ref{spectral_analysis} is representative 
along the entire line of sight, we can derive a range of local ISM densities by dividing by 
the distance limits. For a column density of $N_\mathrm{H} = 6 \times 10^{20}$\,cm$^{-2}$,
and distances of 0.45--1.2\,kpc, the resulting local density $n_\mathrm{H} = 0.42$--$0.16$\,cm$^{-3}$. 
Inputting these into the SNR evolutionary model calculator provided by \cite{2017AJ....153..239L}, 
with otherwise standard model and input values, we calculate the range of possible ages as
21--150\,kyr. However, the morphology of the SNR suggests a much lower age, and therefore we 
suggest the SNR is likely to be at the closer, younger, and higher $n_\mathrm{H}$ ends of 
the allowable ranges.

Taking into account the fact that no pulsar has been associated with the object  so far, it is highly
possible that Hoinga is the remnant of a type Ia SN. This would also be consistent with the high 
latitude of the SNR, as the massive star progenitors of core-collapse SNe are expected to be more 
concentrated in the Galactic plane \citep{Taylor1993,Cordes2002,Faucher-Giguere2006}. 

eROSITA will perform a total of eight all-sky surveys. With further surveys completed, more 
data from the Hoinga remnant will become available in the next few years. This will allow us 
to study the remnants fine structure and spectral properties in more detail, hopefully allowing 
us to further constrain its distance, age, chemical composition, and SN type. The findings of 
Hoinga represent a highlight of the beginning of a wider program setup by the authors WB and 
NHW as part of an eROSITA-Australian-based joint-venture collaboration defined to explore 
the X-ray-radio-sky in order to uncover further exciting surprises in the SNR sphere. 

\begin{acknowledgements}
We thank Bernd Aschenbach and Nicholas Pingel for fruitful discussions and the anonymous referee for 
valuable comments.\\

eROSITA is the primary instrument aboard SRG, a joint Russian-German science mission supported by the
Russian Space Agency (Roskosmos), in the interests of the Russian Academy of Sciences 
represented by its Space Research Institute (IKI), and the Deutsches Zentrum für Luft- und Raumfahrt
(DLR). The SRG spacecraft was built by Lavochkin Association (NPOL) and its subcontractors, and is
operated by NPOL with support from IKI and the Max Planck Institute for Extraterrestrial Physics (MPE).
The development and construction of the eROSITA X-ray instrument was led by MPE, with contributions from
the Dr.~Karl Remeis Observatory Bamberg \& ECAP (FAU Erlangen-N\"urnberg), the University of Hamburg 
Observatory, the Leibniz Institute for Astrophysics Potsdam (AIP), and the Institute for Astronomy and
Astrophysics of the University of T\"ubingen, with the support of DLR and the Max Planck Society. 
The Argelander Institute for Astronomy of the University of Bonn and the Ludwig Maximilians Universit\"at 
Munich also participated in the science preparation for eROSITA. The eROSITA data shown here were
processed using the eSASS/NRTA software system developed by the German eROSITA consortium.\\

NHW is supported by an Australian Research Council Future Fellowship (project number FT190100231) 
funded by the Australian Government. This scientific work makes use of the Murchison Radio-astronomy 
Observatory, operated by CSIRO. We acknowledge the Wajarri Yamatji people as the traditional owners 
of the Observatory site. Support for the operation of the MWA is provided by the Australian 
Government (NCRIS), under a contract to Curtin University administered by Astronomy Australia 
Limited. Establishment of the Murchison Radio-astronomy Observatory and the Pawsey Supercomputing 
Centre are initiatives of the Australian Government, with support from the Government of Western 
Australia and the Science and Industry Endowment Fund. We acknowledge the Pawsey Supercomputing 
Centre which is supported by the Western Australian and Australian Governments. Access to Pawsey 
Data Storage Services is governed by a Data Storage and Management Policy (DSMP). ASVO has received 
funding from the Australian Commonwealth Government through the National eResearch Collaboration 
Tools and Resources (NeCTAR) Project, the Australian National Data Service (ANDS), and the 
National Collaborative Research Infrastructure Strategy. This research has made use of NASA’s 
Astrophysics Data System Bibliographic Services. \\

MGFM acknowledges support by the International Max-Planck Research School on Astrophysics at the 
Ludwig-Maximilians University, IMPRS.\\

This work has made use of data from the European Space Agency (ESA) mission
{\it Gaia} (\url{https://www.cosmos.esa.int/gaia}), processed by the {\it Gaia}
Data Processing and Analysis Consortium (DPAC,
\url{https://www.cosmos.esa.int/web/gaia/dpac/consortium}). Funding for the DPAC
has been provided by national institutions, in particular the institutions
participating in the {\it Gaia} Multilateral Agreement.\\

The following 3rd-party software was used in this work:
{\sc aoflagger} and {\sc cotter} \citep{OffriingaRFI}; \textsc{WSClean} \citep{offringa-wsclean-2014,offringa-wsclean-2017}; {\sc Aegean} \citep{Hancock-2018}; {\sc miriad} \citep{Miriad}; {\sc TopCat} \citep{Topcat} \textsc{NumPy}~v1.11.3 \citep{NumPy,harris2020array}; \textsc{AstroPy}~v2.0.6 \citep{Astropy}; \textsc{SciPy}~v0.17.0 \citep{SciPy}, \textsc{Matplotlib}~v1.5.3 \citep{Matplotlib}.
The manuscript was prepared on the web-based \LaTeX~editor, Overleaf.

\end{acknowledgements}

\bibliographystyle{aa}
\bibliography{snrbib}

\end{document}